\documentclass[review,12pt]{elsarticle}

\usepackage{hyperref}

\usepackage{textcomp,booktabs}
\usepackage[usenames,dvipsnames]{color}
\usepackage{colortbl}

\usepackage{amssymb}
\usepackage{graphicx}
\usepackage{booktabs}
\usepackage{tabularx}
\usepackage{array}
\usepackage{lscape}
\usepackage{longtable}
\usepackage{multirow}
\usepackage{amsmath}
\usepackage{colortbl}
\usepackage{epsfig}
\usepackage{epsf}
\usepackage{subfigure}
\usepackage{rotating}
\usepackage{pdflscape}
\usepackage{lscape}
\usepackage{indentfirst}
\usepackage{mathrsfs}
\usepackage{amsbsy}
\biboptions{numbers,sort&compress}

\journal{IEEE Access}

\begin{document}

\begin{frontmatter}

\title{An adaptive parallel processing strategy in complex event processing systems over data streams}

\author[address1]{Fuyuan Xiao\corref{label1}}
\author[address2]{Masayoshi Aritsugi}
\address[address1]{School of Computer and Information Science, Southwest University, \\ No.2 Tiansheng Road, BeiBei District, Chongqing, 400715, China}
\address[address2]{Big Data Science and Technology, Division of Environmental Science, \\ Faculty of Advanced Science and Technology, Kumamoto University, Japan, \\ 2-39-1 Kurokami, Chuo-ku, Kumamoto 860-8555, Japan}
\cortext[label1]{Corresponding author at: School of Computer and Information Science, Southwest University, Chongqing, 400715, China. E-mail: xiaofuyuan@swu.edu.cn}

%
%
%
%

\begin{abstract}
Efficient matching of incoming events of data streams to persistent queries is fundamental to event stream processing systems.
These applications require dealing with high volume and continuous data streams with fast processing time on distributed complex event processing (CEP) systems.
Therefore, a well-managed parallel processing technique is needed for improving the performance of the system.
However, the specific properties of pattern operators in the CEP systems increase the difficulties of the parallel processing problem.
To address these issues, a parallelization model and an adaptive parallel processing strategy are proposed for the complex event processing by introducing a histogram, and utilizing the probability and queue theory.
The proposed strategy can estimate the optimal event splitting policy, which can suit the most recent workloads conditions such that the selected policy has the least expected waiting time for further processing the arriving events.
The proposed strategy can keep the CEP system running fast under the variation of the time window sizes of operators and input rates of streams.
Finally, the utility of our work is demonstrated through the experiments on the StreamBase system.
\end{abstract}


\end{frontmatter}


\section{Introduction}
Recently, there has been an increasing interest in distributed applications which require processing continuously flowing data from geographically distributed sources to achieve timely responses to complex queries, such as data stream processing (DSP) systems~\cite{abadi2005design,abadi2003aurora,balazinska2008fault,chandrasekaran2003telegraphcq,hwang2007cooperative,shah2004highly,preuveneers2016samurai} and complex event processing (CEP) systems~\cite{diao2007sase,demers2007cayuga,chen2014recommendation,zappia2012lightweight,boubeta2014model,liu2011cube,mei2009zstream,wu2006high,macia2016complex,akdere2008plan,agrawal2008efficient,liu2011high}.
Additionally, the CEP systems focus on detecting patterns of information that represent the higher-level events, which are different with the DSP systems that focus on transforming the incoming flow of information~\cite{cugola2012processing,preuveneers2016security,cugola2012low,ilie2011survey}.
Because the CEP system has many advantages, such as expressive rule language and efficient detection model of events, it has been highly concerned in the academic circles and recently in the industry~\cite{sasehttp,cayugahttp,PIPEShttp,coral8http,streambasehttp,oraclehttp,COEhttp}.
In the CEP systems over data streams, events are processed in real-time for all kinds of purposes, such as wireless sensor networks, financial tickers, traffic management, click-stream inspection, and smart hospital~\cite{boubeta2015model4cep,dunkel2011event,terroso2012cooperative,Hu2016liujing,kim2016rm,bruns2015intelligent,ning2016coordinated,gu2012deadline,Liu2016compressedsensoring,ottenwalder2014mcep,xiao2016efficient}. In these application domains, highly-available event stream processing with fast processing time is critical for handling with the real-world events.

As far as we know, many kinds of parallel methods were devised to deal with massive distributed data streams for the DSP systems~\cite{safaei2010parallel,han2008parallelizing,wu2012parallelizing,hirzel2012partition,johnson2008query,liu2005revisiting,chaiken2008scope,upadhyaya2011latency,schneider2012auto,safaei2012dispatching,brenna2009distributed}.
However, due to the differences between the DSP and CEP systems,
most of the parallel methods that exclusively focus on aggregate queries or binary equi-joins in the DSP systems cannot be simply and directly used in the CEP systems that focus on multi-relational non-equi-joins on the time dimension, possibly with temporal ordering constraints, such as sequences (SEQ) operator and conjunctions (AND) operator~\cite{mei2009zstream,akdere2008plan,xiao2013nested,liu2011high}.
Furthermore, the large volume and input rates of data streams are very common in the big data applications~\cite{carney2002monitoring,fuyuan2012economical}.
The increased time window sizes of operators and input rates of streams may cause bottlenecks of the CEP system.
Bottlenecks can slow down the CEP system.
Even worse, they can result in poor quality of query results which have negative effects on the decision-making.

To address these issues, we propose a parallelization model and an adaptive parallel processing strategy, called $APPS$ by introducing histogram, probability theory and queue theory.
The proposed $APPS$ can estimate the optimal event splitting policy which suits the most recent workloads conditions such that the selected policy has the least expected waiting time for further processing the coming events.
Specifically, the CEP system based on the proposed parallelization model can split the input stream into parallel sub-streams to realise a scalable execution of continuous pattern query.
$APPS$ can keep the CEP system operating at high speed even under the variation of time window sizes of the operators and input rates of the streams.
The utility of our work is substantiated through the experiments on the StreamBase~\cite{streambasehttp} system.

The rest of this paper is organized as follows.
Section~\ref{RelatedWork} discusses the related work in terms of the CEP systems.
Section~\ref{Preliminaries} briefly introduces the preliminaries of this paper.
After that, a parallelization model and three event splitting policies are proposed in Section~\ref{Systemmodel}.
Then, an adaptive parallel processing strategy is proposed to estimate and select the optimal event splitting policy for suiting the workloads conditions in Section~\ref{APPS}.
Section~\ref{Experiment} demonstrates the utility of our proposal through the experiments on the StreamBase system.
Finally, conclusions are given in Section~\ref{Conclusion}.

\section{Related work}\label{RelatedWork}
CEP has had an amount of related work in literature which are discussed as below.
Li et al.~\cite{li2010accelerating} utilised a tree-based CEP approach and optimized the algorithm by event grouping.
Wang et al.~\cite{wang2006bridging} leveraged a directed graph to process the complex event over RFID streams.
Jin et al.~\cite{jin2008efficient} utilised Timed Petri-Net to detect the complex event over RFID streams.
Cayuga~\cite{demers2007cayuga} is an expressive and scalable CEP system which can support on-line detection of a large number of complex patterns over event streams.
SASE~\cite{wu2006high} that leverages NFA and AIS is an optimised complex event detection approach to get high performance and scalability.
But the disadvantages of SASE are that it does not support the complex nested patten query and cannot deal with distributed event stream processing.
To overcome the limitations of SASE, some methods were proposed by extending SASE to make it more powerful and efficient.
Agrawal et al.~\cite{agrawal2008efficient} proposed an improved NFA model to support more powerful query ability.
Zhang et al.~\cite{zhang2010recognizing} improved the SASE model to support imprecise timestamps when processing complex events over data streams.
On the other hand, a research work on detecting complex events over probabilistic event streams based on NFA has been proposed.
Xu et al.~\cite{chuanfei2010complex} used a data structure called Chain Instance Queues to detect complex events which can scan probabilistic streams.
Kawashima et al.~\cite{kawashima2010complex} proposed an optimized method which can calculate the probability of the processed compound events and obtain the value of confidence of the complex pattern.
Shen et al.~\cite{shen2008probabilistic} designed a query language which can express Kleene closure patterns defined by users when detecting probabilistic events.
Due to the efficiency to handle uncertainty and fuse data, some math tools such as fuzzy sets, evidence theory, probability and entropy-based method are widely used in decision making~\cite{deng2015Generalized,ning2016uncertainty,lughofer2007evolving,koppen2014maxmin,dengentropy}.
Some researchers have paid their attention to the applications in CEP system based on probability and fuzzy logic~\cite{wang2013complex,ma2010event,vaiyapuri2016probabilistic}.

In conclusion, the traditional centralized CEP architecture limits its developments, because it is hardly robust and scalable due to single point failure or network break.
In addition, some applications are geographically distributed which need to detect complex events from the distributed system.
Therefore, distributed CEP has been considered recently.
Akdere et al.~\cite{akdere2008plan} devised plan-based CEP across distributed sources.
Ku et al.~\cite{ku2008novel} designed an approach for distributed complex event processing for the RFID application.
Compared with our work, these methods do not support distributed pattern operators in the CEP systems.

\section{Preliminaries}\label{Preliminaries}

\subsection{Event model}
An event which represents an instance and an atomic, is an occurrence of interest at a point in time. Basically, events can be classified into primitive events and composite events. A primitive event instance is pre-defined single occurrence of interest that cannot be split into any small events. A composite event instance that occurs over an interval is created by composing primitive events.

\newtheorem{definition}{\bf Definition}
\begin{definition}\label{d1}
A primitive event $e_i$ is typically modeled multi-dimensionally denoted as $e_i$=$e$($e_i.t$, ($e_i.st=e_i.et$), $<a_1$, $\ldots$, $a_m>$), where, for simplicity, we use the subscript $i$ attached to a primitive $e$ to denote the timestamp $i$, $e_i.t$ is event type that describes the essential features of $e_i$, $e_i.st$ is the start time-stamp of $e_i$, $e_i.et$ is the end time-stamp of $e_i$, $<a_1$, $\ldots$, $a_m>$ are other attributes of $e_i$ and the number of attributes in $e(\cdot)$ denotes the dimensions of interest.
\end{definition}

\begin{definition}\label{d2}
Based on Definition~\ref{d1}, a composite event is denoted as $e$=$e$($e.t$, (($e.st = \min\limits_{1 \leq i \leq n}e_i.st) < (e.et = \max\limits_{1 \leq i \leq n}e_i.et$)), $<a_1$, $\ldots$, $a_g>$).
\end{definition}

\subsection{Nested pattern query language}
We introduce the following nested complex event query language for specifying nested pattern queries:

\noindent
PATTERN  (Event Expression: composite event expressed by the nesting of SEQ and AND, which can have negative event type(s), and their combination operators)\\
WHERE  (Qualification: value constraint)\\
WITHIN  (Window: time constraint)

The composite event expression in the PATTERN clause specifies nested pattern queries, which support nests of SEQ and AND that can have negative event type(s), and their combination operators, as explained above.
Sub-expressions denote inner parts of a pattern query expression.
The value constraint in the WHERE clause defines the context for the composite events by imposing predicates on event attributes.
The time constraint in the WITHIN clause describes the time window during the time difference between the first and the last event instances, which is matched by a pattern query that falls within the window.

\subsection{Pattern operators and their formal semantics}
We define the operators that our method is targeting for.
Specific, in this paper we consider the pattern operators as presented in paper~\cite{liu2011cube,liu2011high}.
In the following, $E_i$ denotes an event type.
More details were presented in~\cite{xiao2013nested}.

\begin{definition}\label{d3}
A SEQ operator~\cite{liu2011cube,liu2011high} specifies a particular order according with the start time-stamps in which the event must occur to match the pattern and thus form a composite event:
\begin{align*}
SEQ(E_i, E_j) = \{<e_i, e_j> | (e_i.st < e_j.st) \wedge (e_i.t = E_i) \wedge (e_j.t = E_j)\}.
\end{align*}
\end{definition}

\begin{definition}\label{d4}
An AND operator~\cite{liu2011high} takes the event types as input and events occur within a specified time window without specified time order:
\begin{align*}
AND(E_i, E_j) = \{<e_i, e_j> | (e_i.t = E_i) \wedge (e_j.t = E_j)\}.
\end{align*}
\end{definition}

\begin{figure}[!ht]
\centering
\includegraphics[width=13cm,clip]{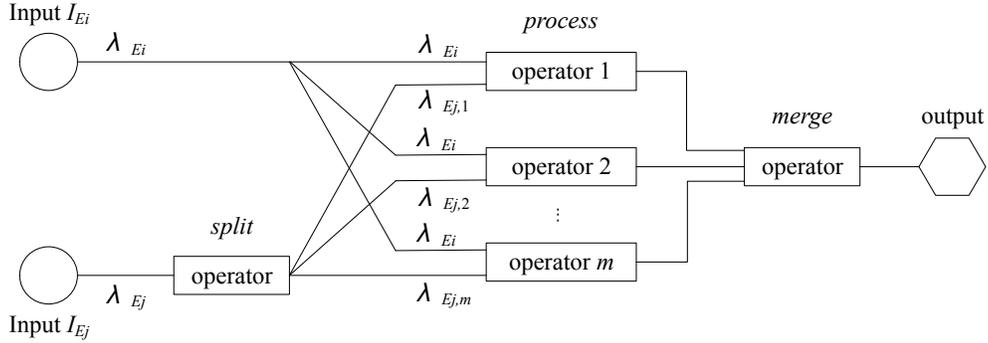}
 \caption{The parallelization model.}\label{parallelismoperator}
\end{figure}

\section{System model}\label{Systemmodel}

\subsection{Parallelization model}\label{Parallmodel}
In this section, we propose a parallelization model that can be utilised for pattern operators which is shown in Fig.~\ref{parallelismoperator}.
We assume that each pattern operator is installed into a server (or host) here.
Because of specific property of pattern operators as described in Section~\ref{Preliminaries}, we can not split both of inputs $I_{E_i}$ and $I_{E_j}$ at the same time.
Otherwise, it will omit detecting some events that may result in wrong decision.
Specifically, once an event of $I_{E_j}$ arrives, the compute function of the pattern operator is initiated.
In other words, the pattern operator creates a new window for every input tuple of $I_{E_j}$.
Therefore, the input stream $I_{E_j}$ is split into parallel sub-streams that will be sent to back-end operators.
The input rate of stream $I_{E_j}$ is equal to the sum of the input rates of sub-streams, i.e., $\lambda_{E_j} = \sum^m_{k=1} \lambda_{{E_j},k}$, where $\lambda_{{E_j},k}$ represents the input rate of sub-stream to the back-end operator $k$.
On the other hand, the replicate of input stream $I_{E_i}$ is directly sent to the back-end operators, each of which has input rate $\lambda_{E_i}$.
We now provide details of the $split-(process*)-merge$ assembly which facilitates the parallelization model of pattern operators.

As shown in Fig.~\ref{parallelismoperator}, the $split-(process*)-merge$ assembly replaces the solo pattern operator in the application data-flow.
In the parallelized version of the application data-flow, $\lambda_{E_j} $ is split to the back-end process operators,
and the output of the pattern operator is replaced by the output coming from the merge operator.

\begin{itemize}
\item
{\bf $split$.} The split operator is to split an input stream into parallel sub-streams. The split operator outputs the incoming events to a number of back-end pattern operators by one of the event splitting policies from Section~\ref{Eventsplittingpolicies} where this selected event splitting policy is estimated by the adaptive parallel processing strategy that will be explained in Section~\ref{APPS}.

\item
{\bf $process$.} The process operator performs the events from the output of the front-end operators. The multiple process operators with the same function can be executing in parallel.

\item
{\bf $merge$.} The merge operator consumes the output events from the process operators to generate the final output events. The merge operator by default simply forwards the output events to its output port.
\end{itemize}

\subsection{Event splitting policies}\label{Eventsplittingpolicies}
In this section, the event splitting policies are given which can be utilised for processing pattern operators in parallel.

\begin{itemize}
\item
{\bf Round-Robin $(RR)$}

Events are assigned to the servers in a cyclical fashion which means that the incoming events will be sent to the downstream servers with equal probability.
This policy equalises the expected number of events at each server.

\item
{\bf Join-the-Shortest-Queue $(JSQ)$}

For the expected number of events, they are assigned to the downstream server with the shortest queue length for further processing.
Here, shortest queue means the queue with the fewest events.

\item
{\bf Least-Loaded-Server-First $(LLSF)$}

For the expected number of events, it dynamically assigns them to the downstream server with the least load.
The least loaded server is the server with the least used memory.

\end{itemize}

\begin{table}
\centering
\caption{Notation}\label{Notation}
\begin{tabular}{|l|l|}\hline
\bf Notation                   &\    \bf Meaning                                             \\ \hline
$\mathcal{P}_j$                &\    event splitting policy $j$                              \\ \hline
$\rho$                         &\    the expected server utilization                          \\ \hline
$\delta$                       &\    threshold of the expected server utilization            \\ \hline
$m$                            &\    degree of parallelization of servers                    \\ \hline
$\mu$                          &\    number of events served per unit time                  \\ \hline
$\lambda_{E_j}$                &\    input rate of input stream $I_{E_j}$                          \\ \hline
$\mathcal{S}_{\nu}$            &\    the $\nu^{th}$ segment of input stream $I_{E_j}$        \\ \hline
$\mathcal{B}_g$                &\    the $g^{th}$ batch partition of a segment               \\ \hline
$i$                            &\    number of events of a batch partition                   \\ \hline
$q$                            &\    number of batch partitions of a segment \\ \hline
$\mathcal{\overline{T}}^{i}_{ps}$   &\    average time devoted to processing $i$ number of events                        \\\hline
$\mathcal{\overline{T}}^{i+1}_{rd}$ &\    average time devoted to re-directing the $(i+1)^{th}$ event                    \\
                                    &\    among servers                                                               \\\hline
$\mathcal{\overline{T}}^{\mathcal{S}}_{ps}$ &\    average time devoted to processing segments                                     \\ \hline
$\mathcal{\overline{T}}^{i}_{es}$   &\    average estimation time devoted for $i$ number of events                      \\ \hline
$\mathcal{T}^{\mathcal{P}_j}_{es}$ &\    estimation time devoted to obtaining optimal $\mathcal{P}_j$ for $\mathcal{S}_{\nu}$  \\ \hline
$E[W^R_i]$ &\    expected redirect time for the events at host $i$ \\ \hline
$E[W^H_i]$ &\    expected waiting time for the events at host $i$ \\ \hline
$E[W_{\mathcal{P}_j}]$ &\    expected waiting time for policy $\mathcal{P}_j$ \\ \hline
\end{tabular}
\end{table}

\section{Adaptive parallel processing strategy}\label{APPS}
\begin{figure*}[hp!]
\centering
\includegraphics[width=19cm,angle=90,clip]{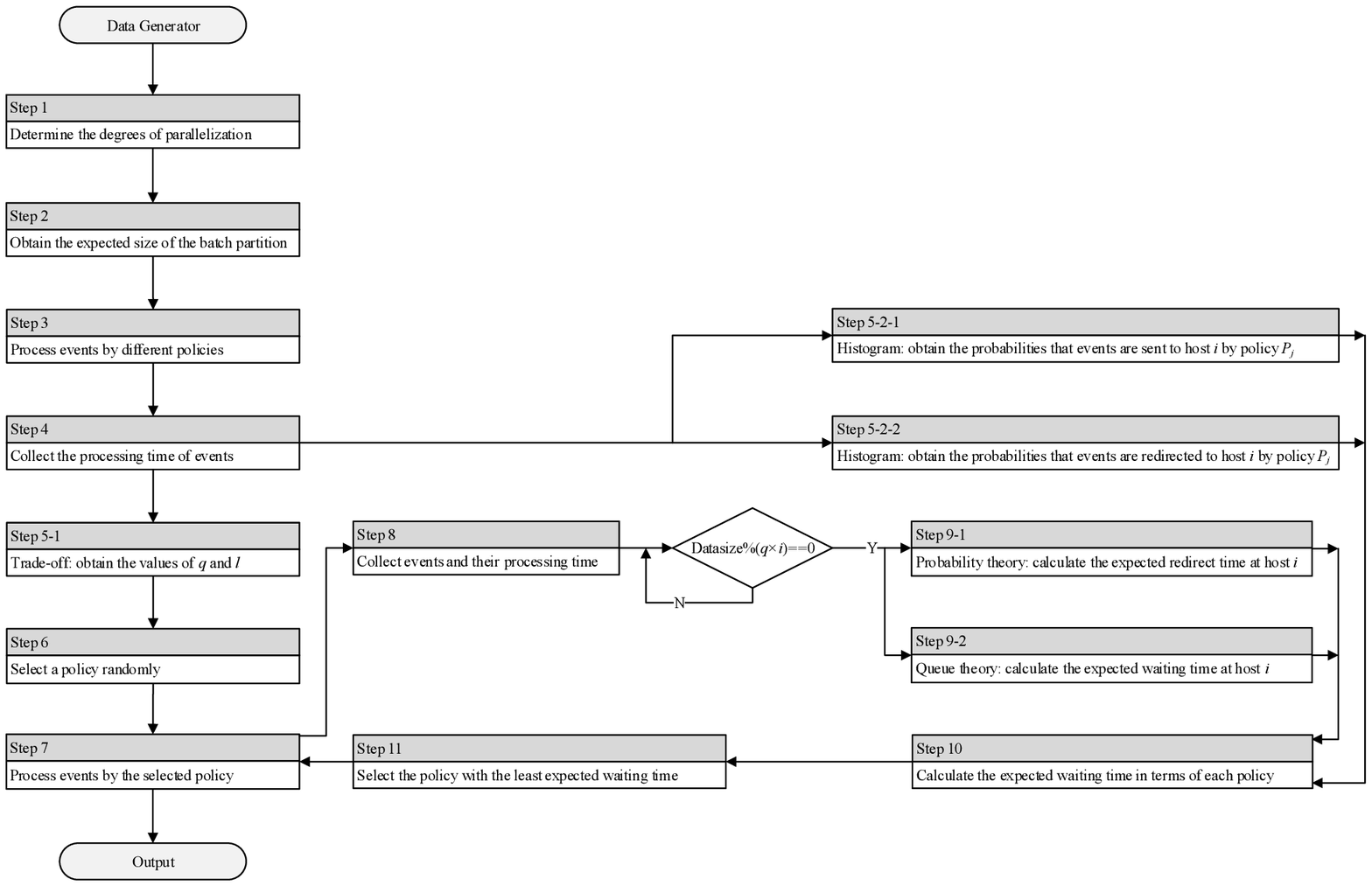}
\caption{The flowchart of the adaptive parallel processing strategy.}
\label{APPSflowchart}
\end{figure*}

In this section, an adaptive parallel processing strategy $(APPS)$ is proposed to estimate and select the optimal event splitting policy which can suit the most recent workloads conditions such that the selected policy is with the least expected waiting time for processing the coming events.
Table~\ref{Notation} shows the key notations that are used in the remainder of this paper.
Fig.~\ref{APPSflowchart} describes the flowchart of the adaptive parallel processing strategy.

\subsection{Degrees of parallelization}
The aim of this stage is to decide the degrees of parallelization for pattern operators in CEP system to be used for processing data streams.

Let $\rho$ be the expected server utilization, $\mu$ be the service rate, $m$ be the number of servers, and $\delta$ be the threshold of the expected server utilization that can be defined by the system administrator in advance according to the implication requirement. By applying queueing theory~\cite{dattatreya2008performance}, $\rho$ is given by

\begin{equation}\label{parallelizationDegrees}
\begin{aligned}
&\rho = \frac{\lambda}{m \mu}, \\
s.t. \quad &\rho \leq \delta, \quad 0 < \delta \leq 1.
\end{aligned}
\end{equation}

Based on~Eq(\ref{parallelizationDegrees}), we can obtain the degrees of parallelization for the pattern operator, i.e., the number of processing servers is as follows:
\begin{equation}\label{servernumber}
m \geq \frac{\lambda}{\mu \delta}, \quad 0 < \delta \leq 1; m \in N.
\end{equation}

\subsection{Expected size of the batch partition}\label{Exp-size-batch}
For further parallel processing, the input stream $I_{E_j}$ needs to be divided into batch partitions.
Because the number of events within each batch partition of segment $\mathcal{S}_{\nu}$ of input stream $I_{E_j}$ should not exceed the threshold of the expected utilization of a single server, the number of events of a batch partition $i$ should satisfy the following condition:

\begin{equation}\label{ExpectedsizeofBP}
\begin{aligned}
 &&i = \mu \delta, \quad i \in N.
\end{aligned}
\end{equation}

\subsection{Event processing time collection}
The aim of this stage is to collect the processing time of the events from the last event type matched by the pattern operator, which are used in the on-line estimation step to estimate various distributional properties of the processing time distribution.

For each new event arriving at the split operator, it records the arrival time of the event.
These values are stored within the event.
The arrived events of a segment of the input stream are then assigned to a back-end server by using the estimated policy $\mathcal{P}_j$, where $\mathcal{P}_j$ denotes the event splitting policy with the least expected waiting time to process the arrived events.
Further details about how to select an appropriate event splitting policy on-line is discussed in Section~\ref{policies}.
For each event that completes processing, its departure time will be stored at the assigned server.
Next, the arrival time and departure time of the event will be sent to its back-end operator.
Then, its corresponding processing time will be calculated by subtracting the departure time, which contributes to the last output event matched by the pattern operator that falls within the time window, from the arrival time recorded by the split operator.

\begin{figure*}[hp!]
\centering
\includegraphics[width=18.5cm,angle=90,clip]{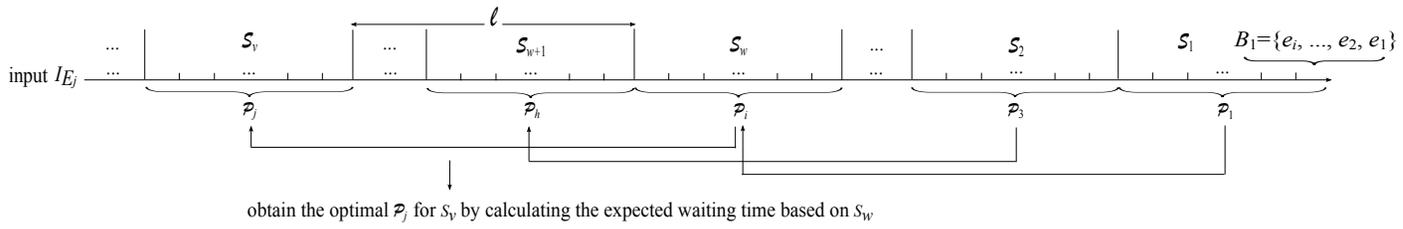}
\caption{Example of obtaining an appropriate policy for processing the coming events.}
\label{BatchPartition}
\end{figure*}

\subsection{Trade-off between the estimation accuracy and the processing time}
Fig.~\ref{BatchPartition} depicts an example of obtaining an appropriate policy for processing the further coming events.
$\mathcal{S}_{\omega}$ denotes the $\omega^{th}$ segment of input stream $I_{E_{j}}$.
$\mathcal{B}_1$ represents the $1^{st}$ batch partition of the segment which consists of events $\{e_1, e_2, \ldots, e_i\}$.
The policy $\mathcal{P}_j$ under segment $\mathcal{S}_{\nu}$ means these events in $\mathcal{S}_{\nu}$ will be processed by using $\mathcal{P}_j$ in which this estimated $\mathcal{P}_j$ is selected based on the empirical data $\mathcal{S}_{\omega}$.
Therefore, we can notice that the time devoted to processing previous $\ell$ number of segments over $m$ parallel servers should exceed the time devoted to estimating an appropriate policy $\mathcal{P}_j$ for segment $\mathcal{S}_{\nu}$.
Otherwise, it introduces extra delay due to waiting for obtaining the optimal policy.
In addition, for obtaining the most accurate expected processing time for $\mathcal{S}_{\nu}$, mean squared error is considered.

Consequently, we treat the accuracy of estimation and the processing time of segments as an integrate constrained optimisation problem.
One objective ($O_1$) tries to maximise the accuracy of estimation, namely, minimising the mean squared error of estimation.
On the other hand, the other objective ($O_2$) tries to maximise the processing time of segments to avoid introducing extra delay for selecting the optimal policy.
Due to the conflicting nature of the different objectives, we obtain the solution by integrating them into one objective and the optimisation problem can thus be formulated as:

\begin{equation}\label{integratedobjective}
\begin{aligned}
&\min {\frac{O_1}{O_2}}.
\end{aligned}
\end{equation}

On the basis of statement~(\ref{integratedobjective}), the values of $q$ and $\ell$ can be obtained and to be used in further on-line event splitting policy selection procedure in Section~\ref{policies}.

\textbf{Objective $O_1$: Mean squared error of estimation constraint.}

In this paper, a general expression is derived for the expected waiting time by applying queueing theory~\cite{newell2013applications}, denoted as $f (E[W])$.
Let $\hat{f}_{\mathcal{S}_{\omega}} (E[W])$ be the expected processing time of the events of segment $\mathcal{S}_{\nu}$ in terms of empirical data $\mathcal{S}_{\omega}$.
Then, $\hat{f}_{\mathcal{S}_{\omega}} (E[W])$ is compared with $f_{\mathcal{S}_{(\nu-1)}} (E[W])$ by the following mean squared error ($MSE$):

\begin{equation}\label{O_1}
\begin{aligned}
&MSE = \frac{1}{\frac{\tau}{q}-\ell-1} \sum^{\frac{\tau}{q}}_{\omega=\nu-1-\ell,\atop \nu=\ell+2} (\hat{f}_{\mathcal{S}_{\omega}} (E[W]) - f_{\mathcal{S}_{(\nu-1)}} (E[W]))^2, \\
with \quad &f (E[W]) =  \frac{\rho^{\sqrt{2(m+1)}-1}}{\mu m (1-\rho)} (\frac{C^2_a + C^2_s}{2}),\quad  \\
&1 \leq q, \ell \leq \tau; 1 \leq \omega \leq \frac{\tau}{q}, \\
s.t.   \quad &MSE < \beta,
\end{aligned}
\end{equation}
in which $q$ is the number of batch partitions of segment $\mathcal{S}_{\nu}$,
and $\ell$ is the difference value subtracting ($\nu-1$) of $\mathcal{S}_{\nu-1}$ from $\omega$ of $\mathcal{S}_{\omega}$ which is devoted to estimating policy $\mathcal{P}_j$ for segment $\mathcal{S}_{\nu}$.
$C^2_a$ represents the squared coefficient of variation of inter-arrival times and $C^2_s$ represents the squared coefficient of variation of service times where they can be obtained by testing.
$f_{\mathcal{S}_{(\nu-1)}} (E[W])$ is the true expected processing time of the events of segment $\mathcal{S}_{\nu}$ in terms of empirical data $\mathcal{S}_{(\nu-1)}$, because it is the nearest empirical data of $\mathcal{S}_{\nu}$ for obtaining the most accurate expected processing time.
$\beta$ denotes the threshold of mean squared error of estimation that can be defined by the system administrator in advance according to the implication requirement.

\textbf{Objective $O_2$: Processing time constraint.}

Let $\mathcal{\overline{T}}^{i}_{ps}$ be the time devoted to processing $i$ number of events,
$\mathcal{\overline{T}}^{i+1}_{rd}$ be the time devoted to re-directing the $(i+1)^{th}$ event,
and $\mathcal{\overline{T}}^{i}_{es}$ be the estimate time for $i$ number of events.
The objectives $O_2$ should satisfy the following condition:

\begin{equation}\label{O_2}
\begin{aligned}
&\mathcal{\overline{T}}^{\mathcal{S}}_{ps} = q \mathcal{\overline{T}}^{i}_{ps} + (q-1) \mathcal{\overline{T}}^{i+1}_{rd}, \\
s.t. \quad &\ell \frac{\mathcal{\overline{T}}^{\mathcal{S}}_{ps}}{m} > \mathcal{T}^{\mathcal{P}_j}_{es}, \\
with \quad &\mathcal{T}^{\mathcal{P}_j}_{es} = q \mathcal{\overline{T}}^{i}_{es}, \\
\quad &1 \leq q, \ell \leq \tau.
\end{aligned}
\end{equation}

The values of $\mathcal{\overline{T}}^{i}_{ps}$, $\mathcal{\overline{T}}^{i+1}_{rd}$ and $\mathcal{\overline{T}}^{i}_{es}$ can be obtained via testing.
If the number of events within one batch partition of $\mathcal{S}_{\nu}$ is large enough, while the time for re-directing each batch partition is quite smaller than the time for processing each batch partition, we can omit $\mathcal{\overline{T}}^{i+1}_{rd}$ in Eq~(\ref{O_2}).

\subsection{On-line selection of event splitting policies}\label{policies}
This stage is pretty critical in the proposed adaptive parallel processing strategy which can estimate and decide the appropriate policy on-line.
In order to calculate the expected waiting time for the policies, we first leverage the histogram to obtain the probabilities that the events are sent to host $i$ by policy $\mathcal{P}_j$, denoted as $\mathcal{P}_j^{H_i}$, and the probabilities that the events are redirected to host $i$ by policy $\mathcal{P}_j$, denoted as $\mathcal{P}_j^{R_i}$.

Next, we introduce queue theory to get the expected waiting time for the events at host $i$ which is formulated as:

\begin{equation}\label{expectedwaitingtime}
\begin{aligned}
E[W^H_i] = \frac{1}{\mu_i} (\frac{\rho_i}{1-\rho_i})(\frac{C^2_{ia}+C^2_{is}}{2}),
\end{aligned}
\end{equation}
where $\rho_i$ denotes the expected server utilization at host $i$,
$\mu_i$ represents the number of events served per unit time at host $i$,
$C^2_{ia}$ represents the squared coefficient of variation of inter-arrival times at host $i$,
and $C^2_{is}$ represents the squared coefficient of variation of service times at host $i$ where they can be obtained on-line.

Additionally, we utilise probability theory to calculate the expected redirect time for the events at host $i$ which is formulated as:

\begin{equation}\label{expectedredirecttime}
\begin{aligned}
E[W^R_i] = \sum^{k}_{r=1} x_r f(x_r).
\end{aligned}
\end{equation}

Based upon the probabilities that events are sent and redirected to different hosts, the expected waiting time for the events at different hosts, and the expected redirect time at different hosts, we then calculate the expected waiting time for all the policies in the list of $APPS$.
$APPS$ derives an general expression for the expected waiting time for policy $\mathcal{P}_j$, denoted as $E[W_{\mathcal{P}_j}]$, by applying probability theory to select the event splitting policy with the least expected waiting time which can be formulated as:

\begin{equation}\label{expectedwaitingtimeforpolicy}
\begin{aligned}
E[W_{\mathcal{P}_j}] = \sum^{h}_{i=1} (\mathcal{P}_j^{H_i} E[W^H_i] + \mathcal{P}_j^{R_i} E[W^R_i]). \\
\end{aligned}
\end{equation}

\section{Experimental evaluation}\label{Experiment}
Based on the parallelization model in Fig~\ref{parallelismoperator}, we implemented the experiments on the StreamBase~\cite{streambasehttp} system for Query $q_1$.

\begin{equation*}
\begin{aligned}
q_1:\quad &\text{PATTERN}\ SEQ(E_1, E_2) \\
&\text{WHERE [Id]}\\
&\text{WITHIN 1 s}
\end{aligned}
\end{equation*}

In order to prove the utility and effectiveness of our proposal, we compared the $APPS$ with $RR$, $JSQ$ and $LLSF$ methods.
We ran the experiments on the machines each of which has AMD Opteron(tm) Processor 6376 and 4.00 GB main memory.
Streams used in the experiments were generated synthetically.
We define the processing time as the difference between the departure time, which contributes to the last output event matched by the pattern operator that falls within the time window, and the arrival time recorded by the split operator.
We provided four machines for $APPS$, $RR$, $JSQ$ and $LLSF$: one machine that creates input data and split the input stream into back-end machines, another two machines that are equipped with $SEQ$ operators with the same functions to process the input streams in parallel, and the other machine that receives data and outputs throughput.
Then, we compared the performance of these methods under different parameter setting in terms of input rate and time window size.

\begin{figure}[!ht]
\centering
\includegraphics[width=10cm,clip]{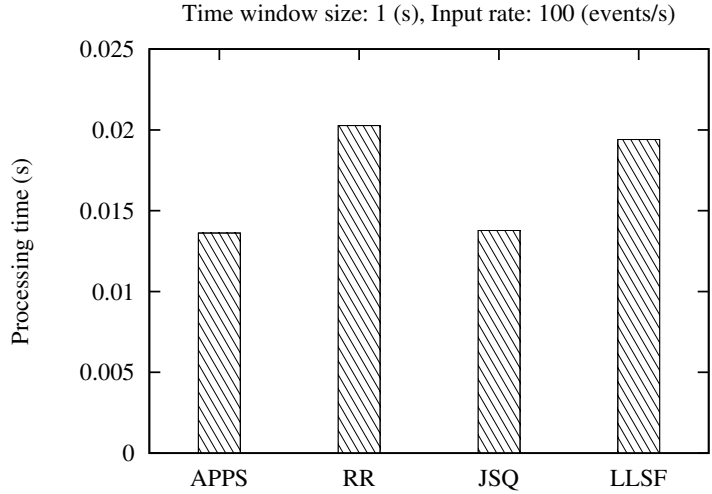}
\caption{Comparing the processing time of the methods.}
\label{processingtime}
\end{figure}

\subsection{Comparing the processing time of the methods}
In this experiment, the input rates were set as 100 events/s, and time window sizes were set as 1 s.
From the experimental result as shown in Fig.~\ref{processingtime}, it was obvious that $APPS$ and $JSQ$ had the lower processing time comparing with $RR$ and $LLSF$ methods.
Because $APPS$ could estimate and select the optimal event splitting policy for further processing the coming events, it had almost the same processing time with $JSQ$ method.

\begin{figure}[!ht]
\centering
\includegraphics[width=10cm,clip]{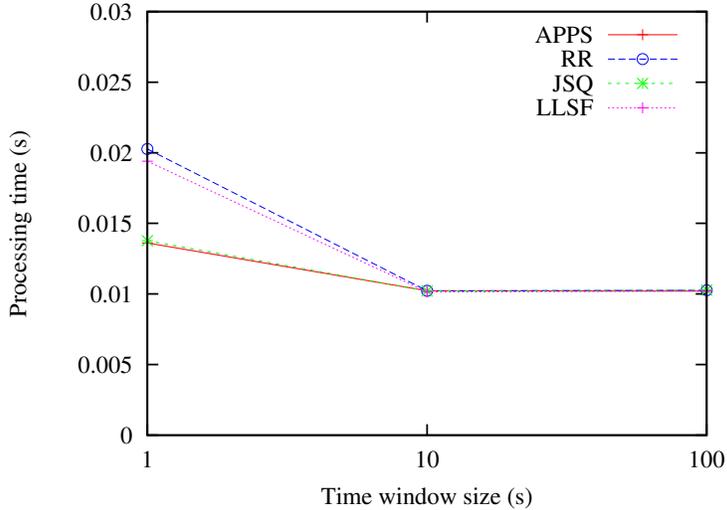}
\caption{Comparing the methods under the variation of time window sizes.}
\label{varyTW}
\end{figure}

\subsection{Varying the time window sizes of operators}
In this experiment, the input rates were set as 100 events/s, while the time window sizes were varying from 1 up to 10, and 100 s.
From the experimental result as shown in Fig.~\ref{varyTW}, we can notice that $APPS$ had almost the same processing time with $JSQ$ method, in which they outperformed $RR$ and $LLSF$ methods, especially when the time windows sizes were set as 1 s.
Whereas, as the time windows sizes increased as 10 up to 100 s, $APPS$, $RR$, $JSQ$ and $LLSF$ methods almost had the same performance.
The reason is that as the time windows sizes increased as 10 up to 100 s, it reached the limitation of processing capacity of machines.

\begin{figure}[!ht]
\centering
\includegraphics[width=10cm,clip]{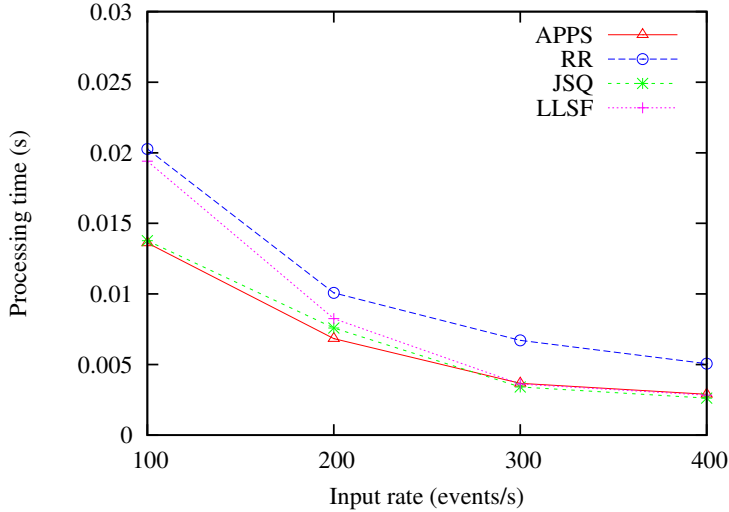}
\caption{Comparing the methods under the variation of input rates.}
\label{varyIR}
\end{figure}

\subsection{Varying the input rates of streams}
In this experiment, the time window sizes were set as 1 s, while the input rates were varying from 100 up to 200, 300, and 400 events/s.
From the experimental result as shown in Fig.~\ref{varyIR}, we can obviously see that the performance of $APPS$ was significantly better than the performance of $RR$, $JSQ$ and $LLSF$ methods.
Because $APPS$ which suits the most recent workloads conditions estimated and selected the optimal event splitting policy for further processing the coming events, it could handle with the input rate variation environment.
On the other hand, as the input rates increased as 100 up to 200, 300, and 400 events/s, $JSQ$ and $LLSF$ methods had the lower processing time than $RR$ method, because $JSQ$ assigned the events to the back-end server with the shortest queue length, while $LLSF$ assigned the events to the back-end server with the least load for further processing the coming events.

\section{Conclusions}\label{Conclusion}
In this paper, we started off with identifying the general problems of adaptive parallel processing with respect to pattern operators in CEP systems.
We proposed a new parallelization model and adaptive parallel processing strategy to estimate the optimal event splitting policy which can suit the most recent workloads conditions such that the selected policy had the least expected waiting time for further processing the coming events.
The utility of our work was demonstrated through the experiments on the StreamBase system.

\section{Acknowledgement}\label{Ack}
This research was partially supported by the Fundamental Research Funds for the Central Universities (Nos. XDJK2015C107, SWU115008),
the Education Teaching Reform Program of Higher Education (No. 2015JY030),
the 1000-Plan of Chongqing by Southwest University (No. SWU116007),
the National Natural Science Foundation of China (Nos. 61672435, 61702427, 61702426),
and the JSPS KAKENHI Grant (No. 15H02705).

%
%

\clearpage



\end{document}